\documentclass[jcp,twocolumn,groupedaddress,preprintnumbers,amsmath,amssymb,4paper,superscriptaddress,floatfix,twoside]{revtex4-1}

\usepackage[yyyymmdd,hhmmss]{datetime}

\usepackage[pdftex]{graphicx}
\usepackage{xcolor}
\usepackage{dcolumn}
\usepackage{bm}
\usepackage{amsmath}

\usepackage{natbib}
\usepackage{hyperref} 
\usepackage{times}

\begin{document}

\sloppy 

\title{OrthoBoXY: A Simple Way to Compute True Self-Diffusion Coefficients from 
MD Simulations with Periodic Boundary Conditions Without Prior Knowledge of
the Viscosity}

\author{Johanna Busch}
\affiliation{Institut f\"ur Chemie, Abteilung Physikalische und Theoretische Chemie, 
Universit\"at Rostock, Albert-Einstein-Str.\ 27, D-18059 Rostock, Germany}

\author{Dietmar Paschek} 
\email{dietmar.paschek@uni-rostock.de}
\affiliation{Institut f\"ur Chemie, Abteilung Physikalische und Theoretische Chemie, 
Universit\"at Rostock, Albert-Einstein-Str.\ 27, D-18059 Rostock, Germany}

\date{\today~at~\currenttime}

\begin{abstract}
Recently, an analytical expression for the system size dependence and direction-dependence of self-diffusion coefficients for neat liquids due to hydrodynamic interactions 
has been derived for molecular dynamics (MD) simulations using orthorhombic unit cells. 
Based on this description, we show that for systems with a ``magic'' box length ratio of $L_z/L_x\!=\!L_z/L_y\!=\!2.7933596497$ the computed self-diffusion coefficients $D_x$ and $D_y$ in $x$- and $y$-direction become system-size independent and represent the true self-diffusion coefficient $D_0\!=\!(D_x+D_y)/2$. Moreover, by using this particular box geometry, the viscosity can be determined with a reasonable degree of accuracy from the difference of components of the diffusion coefficients in $x$-,$y$- and $z$-direction using the simple expression $\eta\!=\!k_\mathrm{B}T\cdot 8.1711245653/[3\pi L_z(D_{x}+D_{y}-2D_z)]$, where $k_\mathrm{B}$ denotes Boltzmann's constant, and $T$ represents the temperature. MD simulations of TIP4P/2005 water for various system-sizes using both orthorhombic and cubic box geometries are used to test the approach.
\end{abstract}


\maketitle

\section{Introduction}

Self-diffusion coefficients  obtained
from from MD simulations with periodic boundary conditions (PBCs)
show  a systematic system size dependence.\cite{duenweg_1993,yeh_2004}
This effect is
caused by the altered
hydrodynamic interactions 
between particles in a periodic system.\cite{duenweg_1993,yeh_2004,kikugawa_2015,voegele_2016,moultos_2016}
It has been demonstrated for 
simulations of 
polymers in solution \cite{duenweg_1993}, 
TIP3P model water molecules, 
and Lennard-Jones particles \cite{yeh_2004}, 
as well as CO$_2$, n-alkanes, and poly(ethylene glycol) dimethyl ethers 
for a wide variety of conditions \cite{moultos_2016}.
An exact expression,
often referred to as Yeh-Hummer approach,
 has been derived to
describe the effect for simulations with a cubic unit cell as \cite{duenweg_1993,yeh_2004}
\begin{equation}
\label{eq:Dpbc1}
D_0 = D_\text{PBC}+\frac{k_\mathrm{B}T\zeta}{6\pi\eta L}\;,
\end{equation}
with the box size $L$, and the shear viscosity $\eta$.
Here, $D_\text{PBC}$ is the self-diffusion coefficient obtained for
a system with PBCs, and $D_0$ is the self-diffusion coefficient obtained
for $L\rightarrow\infty$. The parameter $\zeta\!\approx\!2.8372974795$ 
is the analogue to a Madelung constant 
\cite{beenacker_1986}
of a cubic lattice, which can be
computed via Ewald summation \cite{beenacker_1986,hasimoto_1959,kikugawa_2015} according to
\begin{eqnarray}
\label{eq:Ewald_iso}
\zeta&=&-L\cdot\Biggl\{
\,\left[\sum_{\mathbf{n}\neq 0}
\frac{\mathrm{erfc}({\alpha}\, n)}{n}\right] +\\\nonumber
&&\frac{\pi}{V}
\left[\sum_{\mathbf{k}\neq 0}
\frac{4 \,e^{-k^2/(4{\alpha}^2)}}{k^2}\right]
-\frac{\pi}{{\alpha}^2V}
-\frac{2{\alpha}}{\sqrt{\pi}}
\Biggr\}
\end{eqnarray}
where $\alpha$ is the Ewald convergence parameter.
The vectors
$\mathbf{n}\!=\!(n_x,n_y,n_z)$,
and
$\mathbf{k}\!=\!(k_x,k_y,k_z)$
are the real  and reciprocal lattice vectors with
$n_{i}\!=\!L m_{i}$ and
$k_{i}\!=\!2\pi\cdot m_{i}/L$
with $m_{i}$ being integer numbers, and
 $n=|\mathbf{n}|$ and $k^2=|\mathbf{k}|^2$, respectively.
Equation \ref{eq:Dpbc1}
has been widely applied to determine the
system-size independent {\em true} self-diffusion coefficient
from MD simulations with PBCs.\cite{Maginn_2019}
However, prior knowledge of the shear viscosity $\eta$ is 
required to perform the correction.

For orthorhombic box geometries, the presence of
unequal box-lengths leads to 
different system-size dependencies for each of the components 
$D_{ii}$  of the diffusion
tensor $\mathbf{D}$ such that 
the self-diffusion tensor becomes anisotropic
even for an isotropic fluid.
To describe such a behavior,
Kikugawa et al.\ \cite{kikugawa_2015} have derived generalized
versions of Equations \ref{eq:Dpbc1} and \ref{eq:Ewald_iso},
which can be applied to systems with an
orthorhombic geometry using
\begin{eqnarray}
\label{eq:Dpbc_aniso}
D_0 &=& D_{\textrm{PBC},ii}
+
\frac{k_\mathrm{B}T\zeta_{ii}}{6\pi\eta L_{i}}
\end{eqnarray}
with $i\in\{x,y,z\}$. Here, $L_{i}$ are the individual box-lengths
of the orthorhombic unit cell and
$D_{\textrm{PBC},{ii}}$ are the components of the 
self-diffusion
tensor in the system with PBCs.
The $\zeta_{ii}$ represent the
direction-dependent  Madelung constant analogues of the
orthorhombic lattice using
\begin{eqnarray}
\label{eq:Ewald_aniso}
\zeta_{ii}&=&-\frac{3}{2}\,L_{i}\cdot\Biggl\{
\,{\frac{1}{2}}\Biggl[\sum_{\mathbf{n}\neq 0}
\frac{\mathrm{erfc}({\alpha}\, n)}{n}\\\nonumber
&&+\frac{n_{i}^2}{n^2}
\left(
\frac{\mathrm{erfc}({\alpha}\, n)}{n}
+\frac{2{\alpha}}{\sqrt{\pi}} e^{-{\alpha}^2n^2}
\right)
\Biggr] \\\nonumber
&&+\frac{\pi}{V}
\Biggl[\sum_{\mathbf{k}\neq 0}
\frac{4 \,e^{-k^2/(4{\alpha}^2)}}{k^2}\\\nonumber
&&-\frac{k_{i}^2}{{\alpha}^2 k^2}
e^{-k^2/(4{\alpha}^2)}
\left( 
1+\frac{4{\alpha}^2}{k^2}
\right)
\Biggr] \\\nonumber
&&
-\frac{\pi}{{\alpha}^2V}
-\frac{\alpha}{\sqrt{\pi}}
\Biggr\}
\end{eqnarray}
with $\mathbf{n}\!=\!(n_x,n_y,n_z)$,
and
$\mathbf{k}\!=\!(k_x,k_y,k_z)$
being real  and reciprocal lattice vectors with
$n_{i}\!=\!L_{i} m_{i}$ and
$k_{i}\!=\!2\pi\cdot m_{i}/L_{i}$,
based on integer numbers for $m_{i}$. Again, we use
$n=|\mathbf{n}|$ and $k^2=|\mathbf{k}|^2$, while
$\alpha$ represents the Ewald convergence parameter.
V\"ogele and Hummer \cite{voegele_2016} have derived a similar expression
using Beenakker's expression for the Rotne-Prager tensor under PBCs.\cite{beenacker_1986}

\section{The {``OrthoBoXY''} Method}

From Equation \ref{eq:Dpbc1} follows that
for a cubic unit cell, the obtained self-diffusion coefficients
$D_\textrm{PBC}$
are always smaller than the true self-diffusion coefficient $D_0$.
For orthorhombic unit cells, however, this does not necessarily need to
be the case.\cite{kikugawa_2015a} In fact, for a unit
cell with $L_x=L_y\neq L_z$, 
diffusion
in $x$- and $y$-direction can even become accelerated 
for certain ratios $L_z/L_x\!=\!L_z/L_y$.\cite{kikugawa_2015a,kikugawa_2015}
Using Equation \ref{eq:Ewald_aniso}, we have determined the exact
ratio where this change in sign occurs: by numerically computing the Madelung constant
analogues
$\zeta_{xx}$, $\zeta_{yy}$, and $\zeta_{zz}$
from Equation \ref{eq:Ewald_aniso}, we have
obtained, in accordance with the analysis of
Kikugawa et al.\ \cite{kikugawa_2015}, 
the condition $\zeta_{xx}\!=\!\zeta_{yy}\!=\!0$ to
be related to
a box geometry with a ``magic'' box-length ratio of
$L_z/L_x\!=\!L_z/L_y\approx 2.7933596497$. 
Since the computation has been performed numerically,
we have determined
$\zeta_{xx}=\zeta_{yy}\!<\! 10^{-10}$ using the box geometry indicated above.
For this geometry, we have also computed the Madelung constant analogue
in $z$-direction to be $\zeta_{zz}\approx 8.1711245653$.
The computations of Equation \ref{eq:Ewald_aniso} and
Equation \ref{eq:Ewald_iso}
discussed above
were performed using double precision floating point arithmetic,
and an Ewald convergence parameter of
${\alpha}\!=\!L_x^{-1}\!=\!L_y^{-1}$
for Equation \ref{eq:Ewald_aniso}
 and
${\alpha}\!=\!L^{-1}$ for Equation \ref{eq:Ewald_iso},
with $m_{i}$ ranging between
$-m_\mathrm{max}\leq m_{i}\leq m_\mathrm{max}$ using
$m_\mathrm{max}\!=\!100$
for both the real and reciprocal lattice summation,
ensuring that the calculations are converged.
\begin{table*}[t]
\caption{\label{tab:md_ortho}
Parameters describing the MD simulations
using an orthorhombic unit cell with
$L_z/L_x=L_z/L_y\approx 2.7933596497$ 
 performed
under NVT and NPT conditions at a temperature of $T\!=\!298\,\mbox{K}$
and a density of
 $\rho\!=\!0.9972\,\text{g}\,\text{cm}^{-3}$ (NVT),
 or a pressure of $P\!=\!1\,\mbox{bar}$ (NPT)
 with
 $N$ indicating the number of water molecules and
$L_x$, $L_y$, and $L_z$ representing the box lengths of the orthorhombic unit cell.  
The direction-dependent self-diffusion coefficients
$D_{\text{PBC},{ii}}$ are 
determined from the slope of the center-of-mass mean square displacement of the 
water molecules.
The true self-diffusion coefficient
$D_0$ is determined according to Equation \ref{eq:d0} and
the shear viscosity $\eta$ is determined according to Equation \ref{eq:eta}. 
The errors indicate a range of $\pm 1\sigma$.
}
\setlength{\tabcolsep}{0.38cm}
        \centering              
\begin{tabular}{cccccc}
\hline\hline\\[-0.2em]
$N$ & 
$L_x, L_y/\text{nm}$ &
$L_z/\text{nm}$ &
$D_0  / 10^{-9}\text{m}^2\text{s}^{-1}$ &
$D_{\mathrm{PBC},zz}  / 10^{-9}\text{m}^2\text{s}^{-1}$ &
$\eta / \text{mPa}\,\text{s}$
\\\hline\\
\multicolumn{6}{c}{NVT:}\\[0.6em]
768  & 2.02050 & 5.64398 & $2.267 \pm 0.039$ & $1.922\pm0.020$ & $0.916\pm0.116$\\
1536 & 2.54566 &7.11097  & $2.283 \pm 0.027$ & $1.989\pm0.010$ & $0.853\pm0.084$\\
3072 & 3.20734 &8.95925  & $2.270 \pm 0.021$ & $2.066\pm0.011$ & $0.975\pm0.113$\\
6144 & 4.04100 &11.28796 & $2.289 \pm 0.019$ & $2.104\pm0.008$ & $0.854\pm0.095$\\[0.6em]
\multicolumn{6}{c}{NPT:}\\[0.6em]
3072 & $3.20734\pm 7.7\times 10^{-5}$ & $8.95924\pm 2.2\times 10^{-4}$  & $2.290 \pm 0.030$ & $2.065\pm0.010$ & $0.884\pm0.124$
\\[0.6em]
\hline\hline
\end{tabular}
\end{table*}

Given that we have two unknowns, $D_0$ and $\eta$, and
three equations, it is always  possible to determine
both $D_0$ and $\eta$ from 
direction-dependent diffusion coefficients obtained from
a single MD simulation run
based on an orthorhombic unit cell. 
However, utilizing MD simulations of an orthorhombic box
with $L_z/L_x=L_z/L_y\approx 2.7933596497$ is
particularly intriguing, since now the 
$x$- and $y$- component of the diffusion tensor become system-size
independent such that
$D_{\textrm{PBC},xx}\!=\!D_{\textrm{PBC},yy}\!=\!D_0$.
Note that for such a case a prior knowledge of the 
shear viscosity is not required
for determining $D_0$, and
the self-diffusion coefficient for an infinitely large system
can be simply obtained via
\begin{equation}
\label{eq:d0}
D_0=\frac{D_{\textrm{PBC},xx}+D_{\textrm{PBC},yy}}{2}\;.
\end{equation}
In fact, 
from Equation \ref{eq:Dpbc_aniso} follows, that
for this case the
shear viscosity can also be computed directly
from the knowledge of the three components of
the diffusion tensor using
\begin{equation}
\label{eq:eta}
\eta= \frac{k_\mathrm{B}T\zeta_{zz}}{3\pi L_z
(D_{\textrm{PBC},xx}+D_{\textrm{PBC},yy}-2D_{\textrm{PBC},zz})}\;
\end{equation}
with 
$\zeta_{zz}\approx 8.1711245653$.  
Moreover, Equation \ref{eq:eta} suggests that 
it is perhaps beneficial
to employ particularly small system sizes for determining 
$\eta$ due to an increasing difference between 
$D_0$ and $D_{\textrm{PBC},zz}$ with decreasing system size.
This approach might therefore offer the opportunity for
determining the viscosity and true self-diffusion coefficient from computationally 
expensive calculations such as
{\em ab initio} MD simulations.
\begin{table}[t]
\caption{\label{tab:md_cubic}
Parameters describing the MD simulations
using a cubic unit cell
 performed
under NVT conditions at a temperature of $T\!=\!298\,\mbox{K}$
and a density of
 $\rho\!=\!0.9972\,\text{g}\,\text{cm}^{-3}$ with
 $N$ indicating the number of water molecules and
$L$ representing box length.  
The self-diffusion coefficients
$D_\text{PBC}$ are
determined from the slope of the center-of-mass mean square displacement of the 
water molecules.
The true self-diffusion coefficient
$D_0$ is obtained for systems with periodic boundary conditions
according to Equation \ref{eq:Dpbc1}
using a shear viscosity of 
$\eta\!=\!(0.900\pm0.051)\,\mbox{mPa\,s}$.
The errors indicate a range of $\pm 1\sigma$.
}
\setlength{\tabcolsep}{0.31cm}
        \centering              
\begin{tabular}{cccc}
\hline\hline\\[-0.2em]
$N$ & 
$L/\text{nm}$ &
$D_\text{PBC}/10^{-9}\text{m}^2\text{s}^{-1}$ & 
$D_0/10^{-9}\text{m}^2\text{s}^{-1}$ 
\\\hline\\[-0.6em]
256  & 1.97300 & $1.932\pm 0.021$    & $2.275\pm 0.027$  \\
512  & 2.48582 & $2.001\pm 0.013$    & $2.273\pm 0.019$  \\
1024 & 3.13194 & $2.068\pm 0.012$    & $2.284\pm 0.016$  \\
2048 & 3.94600 & $2.110\pm 0.010$    & $2.282\pm 0.013$  \\[0.6em]
\hline\hline
\end{tabular}
\end{table}
\begin{figure}
        \includegraphics[width=0.45\textwidth]{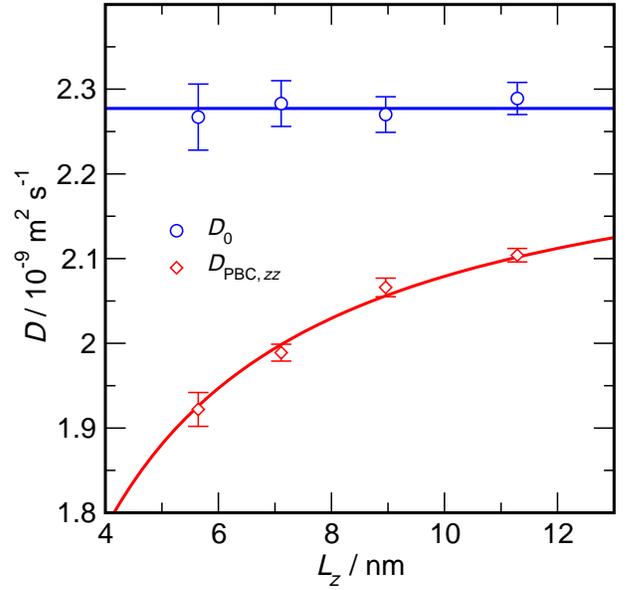}              
        \caption{\label{fig:dplot} Self-diffusion coeffiecients
        of TIP4P/2005 water at $298\,\mbox{K}$ determined from
        MD simulations employing orthorhombic simulation boxes
        with $L_z/L_x\!=\!L_z/L_y\approx 2.7933596497$ for varying
        system sizes. Blue symbols indicate the
        system-size independent $D_0$, which are determined according to 
        Equation \ref{eq:d0}. The blue line indicates the average over all
        system sizes with $D_0=2.277\times 10^{-9}\mbox{m}^2\mbox{s}^{-1}$.
        Red symbols indicate the system-size dependent diffusion coefficient
        in $z$-direction $D_{\textrm{PBC},zz}$. The red line is determined
        according to Equation \ref{eq:Dpbc_aniso} employing
        the average values for $D_0=2.277\times 10^{-9}\mbox{m}^2\mbox{s}^{-1}$
        and $\eta=0.900\,\mbox{mPa\,s}$.
        }
\end{figure}

\section{Molecular Dynamics Simulations}

To test the above outlined 
{\sc OrthoBoXY}
approach, MD  simulations of
TIP4P/2005 model 
water \cite{abascal_2005}
were  carried out, which has been demonstrated to accurately describe the
properties of water compared to other
simple rigid nonpolarizable water models.\cite{vega_2011}.
Simulations were performed
at a temperature of $T\!=\!298\,\mbox{K}$
under NVT  and NPT condititions, either
at a density of
 $\rho\!=\!0.9972\,\text{g}\,\text{cm}^{-3}$ (NVT),
 or at a pressure of
 $P\!=\!1\,\mbox{bar}$ (NPT).
Various system sizes are used for both 
cubic and orthorhombic box geometries.
 MD simulations of 10\,ns length each were performed
 using \textsc{Gromacs} 5.0.6.\cite{gromacs4,gromacs3}
The integration time step for all simulations was $2\,\mbox{fs}$.
The temperature of the simulated systems was controlled employing the
Nos\'e-Hoover thermostat~\cite{Nose:1984,Hoover:1985}
with
a coupling time $\tau_T\!=\!1.0\,\mbox{ps}$.
Constant pressure simulations were realized using
a Rahman-Parrinello-barostat
\cite{Parrinello:1981,Nose:1983} employing $\tau _\textrm{p} = 2.0\,\mbox{ps}$ and $\chi
_\textrm{T}\!=\!33\cdot10^{-6}\,\mbox{bar$^{-1}$}$.
Both, the Lennard-Jones and electrostatic interactions were treated by smooth 
particle mesh Ewald summation.\cite{Essmann:1995,wennberg_2013,wennberg_2015}
The Ewald convergence parameter 
was set to a relative accuracy of the Ewald sum of $10^{-5}$ 
for the Coulomb-interaction and $10^{-3}$ for the LJ-interaction. 
All bond lengths were kept fixed during the simulation run and
distance constraints were solved by means of 
the SETTLE procedure. \cite{miyamoto_1992}
The simulations were carried out in 20 subsequent segments of 
$500\,\mbox{ps}$ length. All reported properties were 
then calculated for 
those segments separately in order to be able to estimate the 
error
using standard statistical analysis procedures.\cite{allentildesley,numrecipes}

\section{Results and Discussion}

Self-diffusion coefficients were computed 
from the slope of the center-of-mass mean square displacement of the 
water molecules using the Einstein formula \cite{allentildesley} according to
\begin{equation}
D_\textrm{PBC}
=\frac{1}{6}
\frac{\partial}{\partial t}
\lim_{t\rightarrow\infty}
\left<
|\mathbf{r}(0)
-
\mathbf{r}(t)
|^2
\right>\;,
\end{equation}
and
\begin{equation}
D_{\textrm{PBC},{ii}}
=\frac{1}{2}
\frac{\partial}{\partial t}
\lim_{t\rightarrow\infty}
\left<
|r_{i}(0)
-
r_{i}(t)
|^2
\right>\;,
\end{equation}
where $\mathbf{r}(t)=[r_x(t),r_y(t),r_z(t)]$ represent the position of the center of mass
of a water molecule at time $t$ and the $r_{i}(t)$ are its respective components
in $x$-, $y$-, and $z$-direction.
All computed self-diffusion coefficients 
shown Tables \ref{tab:md_ortho} and 
\ref{tab:md_cubic} were
determined from the slope of the mean square
displacement of the water molecules
fitted to time intervals between $15\,\mbox{ps}$ and $200\,\mbox{ps}$.

Table \ref{tab:md_ortho} contains results from MD simulations using
orthorhomic unit cells with
$L_z/L_x=L_z/L_y\approx 2.7933596497$ 
for system sizes between 768 and 6144 water molecules,
while Table \ref{tab:md_cubic} contains the data
obtained for cubic unit cells 
with system sizes between 256 and 2048 water molecules.
The diffusion coefficients $D_0$ obtained from the 
simulations based on an
orthorhombic system, shown
in Figure \ref{fig:dplot} and given 
in Table \ref{tab:md_ortho}, exhibit no systematic system size
dependence. Here the average over the different system sizes
is determined to be
$D_0=(2.277\pm0.013)\times 10^{-9}\,\text{m}^2\text{s}^{-1}$.
As shown in Figure \ref{fig:dplot},
the computed self-diffusion coefficients in 
$z$-direction $D_{\mathrm{PBC},zz}$, however, show a strong system
size dependence. From the knowledge of $D_0$ and 
$D_{\mathrm{PBC},zz}$ we compute the shear viscosity 
$\eta$.
The computed viscosities for all systems considered are
shown in Table \ref{tab:md_ortho}. No
systematic system size dependence is observed,
leading to an average value
of $\eta\!=\!(0.900\pm0.051)\,\mbox{mPa\,s}$
for the viscosity of TIP4P/2005 water at $T\!=\!298\,\mbox{K}$
when
averaging over all systems.
Note that the computed errors
of $\eta$ also do not show any systematic variation 
with the system size although the accuracy
of the computed self-diffusion coefficients decreases with decreasing
system size. Possibly, the increasing difference between $D_0$ and
$D_{\mathrm{PBC},zz}$ with decreasing system size 
is compensating for this loss of accuracy, as anticipated earlier.
The computed average viscosity is close to the experimental
value for water of $0.8928\,\mbox{mPa\,s}$ 
at $298\,\mbox{K}$ given by Harris and Woolf.\cite{harris_2004,harris_2004c}
It is, however, slightly larger than the viscosity value of
$0.855\,\mbox{mPa\,s}$ reported by Gonz\'ales and Abascal
\cite{gonzales_2010}, 
and the value of $0.83\pm0.07\,\mbox{mPa\,s}$ reported by
Tazi et al.\ \cite{tazi_2012} for the TIP4P/2005 model.
We would like to point out that this slightly
enhanced viscosity might be related to the fact
 that we applied the 
PME summation for both the Lennard-Jones interactions
and the Coulomb interactions in our simulations.
Note  that the enhanced viscosity is accompanied by a
similarly reduced diffusivity: when scaling
the diffusion coefficient of
$(2.49\pm 0.06)\times 10^{-9}\,\text{m}^2\text{s}^{-1}$, reported
by Tazi et al.\ \cite{tazi_2012}
(which was also determined by applying the Yeh-Hummer correction) by a
factor of $0.83/0.90$, we end up with a diffusion coefficient 
of $2.296 \times 10^{-9}\,\text{m}^2\text{s}^{-1}$,
which 
matches very well the diffusion coefficient determined here.
Both values are lying close to the experimental value
of $2.3\times 10^{-9}\,\text{m}^2\text{s}^{-1}$ 
at $298\,\mbox{K}$.\cite{krynicki_1978}
The computed viscosities shown in Table \ref{tab:md_ortho}
are estimated with a relative accuracy between $10\,\%$ and
$14\,\%$, which is not a particularly impressive.
However, it is
comparable to the accuracy which is available via the integration
over the stress-tensor auto-correlation function reported
by Tazi et al.\ .\cite{tazi_2012}

The diffusion coefficients $D_\textrm{PBC}$ obtained for the cubic systems shown
in Table \ref{tab:md_cubic} exhibit
the familiar system size dependence \cite{yeh_2004} and are corrected according
to Equation \ref{eq:Dpbc1}
using the average shear viscosity of
$\eta\!=\!(0.900\pm0.051)\,\mbox{mPa\,s}$ discussed above.
Again, the computed $D_0$ show no
systematic system size dependence and are leading to an 
average value of
$D_0=(2.279\pm0.010)\times 10^{-9}\,\text{m}^2\text{s}^{-1}$, which is
consistent with our simulations employing orthorhombic unit cells.

To test whether the outlined procedure is also applicable to MD simulations
performed under NPT conditions, we have conducted an additional constant
 pressure simulation of an orthorhombic system 
 using the ``magic'' box-length ratio of
 $L_z/L_x=L_z/L_y\approx 2.7933596497$
for a system-size of $N\!=\!3072$ water molecules, as shown in Table
\ref{tab:md_ortho}. Here, we have applied an equal scaling of the box-lengths
in the Rahman-Parrinello barostat to keep the box-length ratio fixed.
The computed diffusion coefficient
$D_0\!=\!(2.290 \pm 0.030)\times 10^{-9}\,\text{m}^2\text{s}^{-1}$ 
and viscosity of 
$\eta\!=\!(0.884\pm 0.124)\,\mbox{mPa\,s}$ fall well within the range
of data computed from NVT simulations.

\section{Conclusion}

In conclusion, we would like to point out that with 
the proposed {\sc OrthoBoXY} approach of using
an orthorhombic system 
 with a ``magic'' box-length ratio of
 $L_z/L_x=L_z/L_y\approx 2.7933596497$,
we are able to determine the {\em true} (i.e. system size independent)
self-diffusion coefficient $D_0$ for TIP4P/2005 water without prior knowledge
of the shear viscosity 
from a single MD simulation run
by doing nothing more than just employing a particularly odd shaped
simulation box. The computed values for $D_0$ agree with the values
determined from MD simulations employing cubic unit cells by applying the
widely used Yeh-Hummer correction.
In addition, from the analysis of the diffusion coefficients it is also possible
to derive the shear viscosity with an accuracy, 
comparable to the accuracy which is achieved via the integration
over the stress-tensor auto-correlation function.
Both, the computed self-diffusion coefficient and shear viscosity 
agree nearly quantitatively with the experimentally observed
data for water at $298\,\mbox{K}$.

\section*{Acknowledgements}

We thank the computer center at the University of 
Rostock (ITMZ) for providing and maintaining computational resources. The authors thank
J.K. Philipp for proofreading the manuscript.
 
\section*{Data Availability Statement}

The code of 
\href{https://www.gromacs.org}{GROMACS} is freely available. 
Input parameter and topology files for the MD simulations and the 
code for computing the Madelung constant analogues for cubic and orthorhombic
lattices can be downloaded from GitHub  via
\href{https://github.com/Paschek-Lab/OrthoBoXY/}{https://github.com/Paschek-Lab/OrthoBoXY/}

%\clearpage
%\bibliographystyle{unsrtnat}
%\bibliography{all}
%\printbibliography

\end{document}